# Twitter discussions and emotions about COVID-19 pandemic: a machine learning approach

Jia Xue, PhD
Factor-Inwentash Faculty of Social Work
& Faculty of Information
University of Toronto
jia.xue@utoronto.ca

Junxiang Chen, PhD
School of Medicine
University of Pittsburgh
Juc91@pitt.edu

Ran Hu
Factor-Inwentash Faculty of Social Work
University of Toronto
ranh.hu@mail.utoronto.ca

Chen Chen, PhD
Middleware system research group
University of Toronto
Chenchen@eecg.toronto.edu

Chengda Zheng
Faculty of Information
University of Toronto
chengda.zheng@mail.utoronto.ca

Xiaoqian Liu
Institute of Psychology
Chinese Academy of Sciences
liuxiaoqian@psych.ac.cn

Tingshao Zhu*
Institute of Psychology
Chinese Academy of Sciences
tszhu@psych.ac.cn (T.S.Z.)

**\*Corresponding author**: Tingshao Zhu, Professor, CAS Key Laboratory of Behavioral Science, Institute of Psychology, 16 Lincui Road, Chaoyang District, Beijing 100101, China. tszhu@psych.ac.cn (T.S.Z.)

**Acknowledgements:** Publication of this article is supported by Funding by China Social Science Fund (17AZD041). The data collection is supported by Artificial Intelligence Lab for Justice at University of Toronto, Canada.

# Twitter discussions and emotions about COVID-19 pandemic: a machine learning approach

## Abstract

**Background:** Public response to the COVID-19 pandemic is important to be measured. Twitter data are an important source for the infodemiology study of public response monitoring.

**Objective:** The objective of the study is to examine coronavirus disease (COVID-19) related discussions, concerns, and sentiments that emerged from tweets posted by Twitter users.

**Methods:** We analyze 4 million Twitter messages related to the COVID-19 pandemic using a list of 25 hashtags such as "coronavirus," "COVID-19," "quarantine" from March 1 to April 21 in 2020. We use a machine learning approach, Latent Dirichlet Allocation (LDA), to identify popular unigram, bigrams, salient topics and themes, and sentiments in the collected Tweets.

**Results:** Popular unigrams include "virus," "lockdown," and "quarantine." Popular bigrams include "COVID-19," "stay home," "corona virus," "social distancing," and "new cases." We identify 13 discussion topics and categorize them into five different themes, such as "public health measures to slow the spread of COVID-19," "social stigma associated with COVID-19," "coronavirus news cases and deaths," "COVID-19 in the United States," and "coronavirus cases in the rest of the world. Across all identified topics, the dominant sentiments for the spread of coronavirus are *anticipation* that measures that can be taken, followed by a mixed feeling of trust, anger, and fear for different topics. The public reveals a significant feeling of *fear* when they discuss the coronavirus new cases and deaths than other topics.

**Conclusion:** The study shows that Twitter data and machine learning approaches can be leveraged for infodemiology study by studying the evolving public discussions and sentiments during the COVID-19. As the situation evolves rapidly, several topics are consistently dominant on Twitter, such as "the confirmed cases and death rates," "preventive measures," "health authorities and government policies," "COVID-19 stigma," and "negative psychological reactions (e.g., fear)." death. Real-time monitoring and assessment of the Twitter discussion and concerns can be promising for public health emergency responses and planning. Already emerged pandemic fear, stigma, and mental health concerns may continue to influence public trust when there occurs a second wave of COVID-19 or a new surge of the imminent pandemic.

**Introduction**

Eight million people have been confirmed positive of COVID-19 across 110 countries as of mid-June 2020, and the death toll has reached close to 435,000 [1]. The widespread utilization of social media, such as Twitter, accelerates the process of exchanging information and expressing opinions about public events and health crises [2-5]. COVID-19 is one of the trending topics on Twitter since January 2020 and has continued to be discussed to date. Since quarantine measures have been implemented across most countries (e.g., the Shelter-in-Place order in the United States), people have been increasingly relying on different social media platforms to receive news and express opinions. Twitter data are valuable in revealing public discussions and sentiments to interesting topics, and real-time news updates in global pandemics, such as H1N1 and Ebola [6-9]. Chew and Eysenbach's study shows that Twitter can be used for real-time "infodemiology" studies, a source of opinions for health authorities to respond to public concerns [6]. In the current COVID-19 pandemic, many government officials worldwide are using Twitter, as one of the main communication channels, to regularly share policy updates and news related to COVID-19 to the general public [10].

Since the COVID-19 outbreak, a growing number of studies have collected Twitter data to understand the public responses to, and discussions around COVID-19 [11-16]. For instance, Abd-Alrazaq and colleagues adopt topic modeling and sentiment analysis to understand main discussion themes and sentiments around COVID-19, using Tweets collected between February 2 and March 15, 2020 [11]. Budhwani and Sun compare Tweets discussions before and after March 16, 2020, when President Trump tweeted the "Chinese virus," and find a significant increase use of "Chinese virus" in people's Tweets across many states in the United States [14]. Mackey and colleagues analyze about 3,465 tweets collected between March 2 and 20, 2020, using a topic model to explore users' self-reported experiences with COVID-19 and related symptoms [16]. Ahmed and colleagues conduct social network analysis and content analysis of the collected Tweets between March 27 and April 4, 2020, to understand what may have driven the misinformation that linked 5G towers in the United Kingdom to the COVID-19 pandemic [12]. As conversations on Twitter continue to take place and evolve, it is worth continuing to use tweets as a source of data to track and understand what the salient topics discussed on Twitter in response to the COVID-19 pandemic and track their changes in different time periods are.

To expand the literature on public reactions to the COVID-19, the present study aims to examine the public discourses and emotions about the COVID-19 pandemic by analyzing more than 4 million Tweets collected between March 7 and April 21, 2020.

## Methods

### Research design

We used a purposive sampling approach to collect COVID-19 related Tweets published between March 7 and April 21, 2020. Our Twitter data mining approach followed the pipeline displayed in Figure 1. Data preparation included three steps (1) sampling, (2) data collection, and (3) pre-processing the raw data. The data analysis stage included unsupervised machine learning, sentiment analysis, and qualitative method. The unit of analysis was each message-level tweet. Unsupervised learning is one approach in machine learning, and used to examine data for patterns, and derives a probabilistic clustering based on the text data. We chose unsupervised learning because it is commonly used when existing studies have little observations or insights of the unstructured text data [17]. Since a qualitative approach has challenges analyzing large-scale Twitter data, unsupervised learning allows us to conduct exploratory analyses of large text data in social science research. In the present study, we first employed an unsupervised machine learning approach to identify salient latent topics. Using the topics, we used a qualitative approach to develop themes further, as a qualitative approach allows a deeper dive into the data, such as through manual coding and inductively developing themes based on the latent topics generated by machine learning algorithms.

**Figure 1. Twitter data mining pipeline**

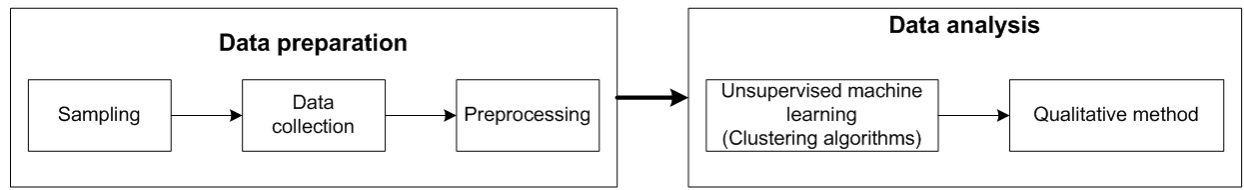

**Sampling and Data collection**

We used a list of COVID-19 related hashtags as search terms to fetch Tweets, such as "#coronavirus," "#2019nCoV," "#COVID19," "#coronaoutbreak," and "#quarantine," (Appendix 1). Twitter's open application programming interface (API) allowed us to collect updated Twitter messages that are set open by default. From March 7 to April 21, 2020, we collected about 35 million (n=35,204,604) Tweets during this period, shown Figure 2. After removing non-English Tweets, 23 million (n=23,817,948) Tweets remained. After removing the duplicates and Retweets (i.e., only reposts the original message without adding any more words), we have more than 4 million (n=4,196,020) Tweets in our final dataset. We collected and downloaded the following features for each Tweet, including (1) Tweet full text, (2) the numbers of favorites, followers, and friends, (3) user' geolocation; and (4) user' description/self-created profile.

**Figure 2. Tweets pre-processing chart**

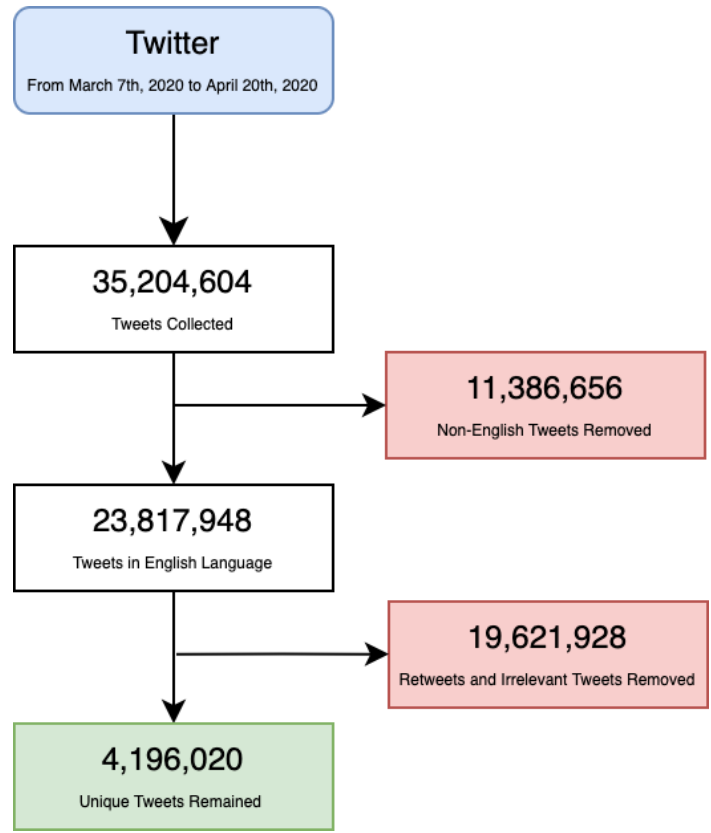

**Pre-processing the raw data**

Shown in Figure 1, we used Python to clean the raw data. The process was as follows [18],

(1) We removed the hashtag symbol, @users, and URLs from the Tweets in the dataset.

(2) We removed non-English characters (non-ASCII characters) because the present study focuses on Tweets in English.

(3) We removed special characters, punctuations, and stop-words listed in [19] from the dataset as they do not contribute to the semantic meanings of the messages.

**Data analysis**

*Unsupervised machine learning*

Latent Dirichlet Allocation (LDA) [20] is one of the widely used unsupervised machine learning approaches allowing researchers to analyze unstructured text data (e.g., Twitter messages). Based on the data itself, the algorithm produces frequently mentioned pairs of words, the pairs of words co-occur together, and the latent topics and their distributions over topics in the document [21]. Existing studies have indicated the feasibility of using LDA in identifying the patterns and themes of the Tweets texts related to COVID-19 [11 22].

*Qualitative analysis*

To triangulate and contextualize findings from the LDA model, we employed a qualitative approach to develop themes further. Specifically, using Braun and Clarke's [23] six steps of thematic analysis: (1) getting familiar with the keyword data, (2) generating initial codes, (3) searching for themes, (4) reviewing potential themes, (5) defining themes, and (6) reporting. Since the thematic approach relies on human interpretation, a process that can be significantly influenced by the personal understanding of the topics and a variety of bias, we had two team members conduct the first five steps independently. Then, the two members reviewed all identified themes and resolved disagreements together. Finally, we finalized themes corresponding to each one of the 13 topics.

*Sentiment analysis*

We used sentiment analysis, a natural language processing (NLP) approach, to classify the main sentiments of a given twitter message, such as fear and joy [24]. In the study, we used the NRC Emotion Lexicon, which consisted of eight primary emotions: anger, anticipation, fear, surprise, sadness, joy, disgust, and trust [25]. We followed the four steps to calculate the emotion index for each twitter message, including (1) removing articles, pronouns (e.g., "and," "the," or "to"), (2) applying a stemmer by removing the predefined list of prefixes and suffixes (e.g., "running" after stemming becomes "run") [26], and (3) calculating the emotion index. We only kept one emotion with the maximum matching counts if one sentence has multiple emotions; and (4) calculating the scores for each eight-emotion type. We discussed the four steps in detail in the previous study [18].

**Results**

**Descriptive results**

A total of four million (n=4,196,020) Tweets consisted of our final dataset after pre-processing all raw data. We identified the most popular tweeted bigrams (pairs of words) related to COVID-19. Bigrams captured "two concessive words regardless of the grammar structure and semantic meaning and may not be self-explanatory" [27], including "covid 19," "stay home," "social distancing," "new cases," "don't know," "confirmed cases," "home order," "New York," "tested positive," "death toll," and "stay safe." Popular unigrams included virus, lockdown, quarantine, people, new, home, like, stay, don't, and cases. We presented the most popular unigrams and bigrams related to COVID-19 in Table 1 and visualized them using the word clouds in figure 3 and figure 4.

Table 1. Top 50 popular bigrams and unigram and their distributions

| Top 50 bigrams | Dataset (%) | Top 50 unigrams | Dataset (%) |
| --- | --- | --- | --- |
| covid 19 | 0.29% | virus | 1.18% |
| stay home | 0.26% | lockdown | 0.98% |
| corona virus | 0.12% | quarantine | 0.94% |
| social distancing | 0.08% | people | 0.82% |
| new cases | 0.07% | coronavirus | 0.79% |
| dont know | 0.04% | new | 0.47% |
| confirmed cases | 0.04% | home | 0.45% |
| home order | 0.04% | like | 0.44% |
| new york | 0.04% | im | 0.41% |
| tested positive | 0.04% | stay | 0.41% |
| death toll | 0.04% | dont | 0.41% |
| home orders | 0.04% | cases | 0.37% |
| quarantine got | 0.03% | time | 0.36% |
| stay safe | 0.03% | covid | 0.35% |
| spread virus | 0.03% | 19 | 0.30% |
| coronavirus cases | 0.03% | need | 0.30% |
| shelter place | 0.03% | day | 0.29% |
| coronavirus pandemic | 0.03% | trump | 0.28% |
| year old | 0.03% | china | 0.28% |
| public health | 0.03% | know | 0.28% |
| chinese virus | 0.03% | going | 0.25% |
| ill deliver | 0.03% | help | 0.25% |
| deliver copy | 0.03% | pandemic | 0.24% |
| health care | 0.03% | world | 0.24% |
| support usps | 0.03% | health | 0.23% |
| signing support | 0.02% | think | 0.22% |
| usps ill | 0.02% | deaths | 0.21% |
| wuhan virus | 0.02% | today | 0.21% |
| quarantine im | 0.02% | good | 0.20% |
| mental health | 0.02% | work | 0.20% |
| dont want | 0.02% | want | 0.19% |
| im going | 0.02% | corona | 0.17% |
| president trump | 0.02% | spread | 0.17% |
| united states | 0.02% | got | 0.17% |
| dont think | 0.02% | support | 0.17% |
| copy officials | 0.02% | government | 0.17% |
| feel like | 0.02% | right | 0.15% |
| looks like | 0.02% | way | 0.15% |
| positive cases | 0.02% | care | 0.15% |
| staying home | 0.02% | social | 0.15% |
| officials toodelivered | 0.02% | news | 0.15% |
| coronavirus outbreak | 0.02% | state | 0.15% |
| domestic violence | 0.02% | country | 0.15% |
| coronavirus lockdown | 0.02% | said | 0.14% |
| healthcare workers | 0.02% | ive | 0.14% |
| people died | 0.02% | days | 0.14% |
| quarantine day | 0.02% | testing | 0.14% |
| donald trump | 0.02% | stop | 0.13% |
| social media | 0.02% | says | 0.13% |

**Figure 3. The word cloud of the most popular unigram**

**Figure 4. The word cloud of the most popular bigrams**

**COVID-19 related topics**

Our approach, Latent Dirichlet Allocation (LDA), produced frequently co-occurred pairs of words related to COVID-19 and organized these co-occurring words into different topics. LDA allowed researchers to manually define the number of topics (e.g., ten topics, twenty topics) that we would like to generate. Consistent with the previous studies, we used the coherence model – gensim [28] to calculate the most appropriate number of topics based on the specific data itself. For this dataset, the number of topics (n=13) returned by LDA had the highest coherence score as well as the smallest topic number. For example, the numbers of topics (n=19 or n=20) had higher coherence scores than the number of topics (n=13), but they represented larger topic numbers, shown in Figure 4.

**Figure 5. The number of topics based on the coherence model**

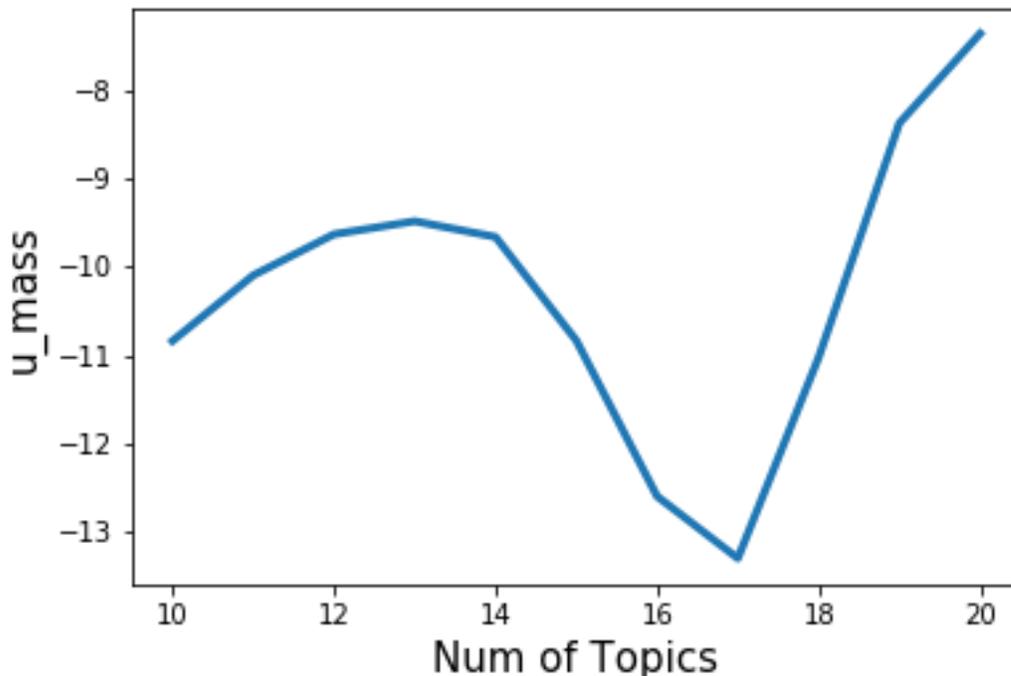

We further analyzed the document-term matrix and obtained the distributions of 13 topics with the chosen thirteen topics. We presented the results of the identified 13 salient topics and the most popular pairs of words within each topic in Table 2. For example, Topic 3 had the highest distribution (8.87%) among all 13 common latent topics. The bigrams were associated with Topic 3, including "tested positive," "coronavirus outbreak," "New York," "shelter place," and "mental health." These pairs of words frequently co-occurred together, and therefore the LDA model assigned them to the same topic.

Table 2. Identified salient topics and their components (bi-grams)

| Topic | Bigrams within topics | Distribution (%) |
|---|---|---|
| 1 | covid 19, dont know, deadly virus, im gonna, spreading virus, 19 lockdown, herd immunity, 000 people, 19 pandemic, dont need, face masks, fox news, health workers, small businesses, home quarantine, like this, virus came, slow spread, test kits, total confirmed | 8.51% |
| 2 | spread virus, health care, staying home, white house, positive cases, people die, 14 days, coronavirus deaths, care workers, ive seen, need help, day lockdown, know virus, im getting, doctors nurses, quarantine period, virus world, stop virus, people getting, week quarantine | 7.24% |
| 3 | tested positive, coronavirus outbreak, wuhan virus, positive coronavirus, confirmed cases, new york, shelter place, mental health, china virus, feel like, new cases, gt gt, coronavirus covid, virus, weeks, people virus, people don't, bringing total, press conference, sars cov | 8.87% |
| 4 | dont think, virus spread, lockdown period, fake news, nursing homes, wuhan lab, best thing, months, lockdown amp, 21 3, id like, people know, real time, entire world, know im, know it, wake up, feel free, dont wanna, anthony fauci | 6.56% |
| 5 | u s, coronavirus cases, public health, save lives, novel coronavirus, long term, south korea, dont forget, bbc news, care homes, news coronavirus, million people, doesnt mean, family members, want know, coronavirus vaccine, going on, rest world, coronavirus, new jersey | 7.36% |
| 6 | at home, stay at, home order, thank you, look like, good news, test positive, people stay, fight virus, people protesting, face mask, good thing, young people, lock down, wearing masks, cases deaths, trump said, deaths reported, shut down, active cases | 7.36% |
| 7 | social distancing, day quarantine, healthcare workers, prime minister, world health, dont care, global pandemic, dont understand, health organization, dr fauci, let know, time lockdown, virus isn't, in place, anti lockdown, shelter in, people think, live updates, 2 months | 7.81% |
| 8 | coronavirus lockdown, coronavirus crisis, amid coronavirus, looks like, new coronavirus, task force, im sure, coronavirus patients, prevent spread, virus doesn't, dont let, long time, new York, high risk, coronavirus task, thank god, number deaths, dont like, virus outbreak, coronavirus cases | 7.47% |
| 9 | stay safe, chinese virus, self quarantine, need know, people going, new virus, common sense, safe stay, virus amp, b c, 2 2, family friends, we've got, got virus, stay away, testing kits, health amp, virus gone, april 20, knew virus | 7.07% |
| 10 | corona virus, new cases, death toll, im going, quarantine day, people died, spread coronavirus, cases coronavirus, people dying, quarantine im, total number, number cases, cases reported, april 2020, confirmed cases, coronavirus death, 24 hours, people need, stop spread | 8.84% |
| 11 | stay home, home orders, president trump, social media, home stay, loved ones, stay safe, death rate, working home, 31 000, social distance, 3100 000, protesting stay, breaking news, deaths, im sorry, 10 000, mortality rate | 8.67% |
| 12 | coronavirus pandemic, year old, united states, wash hands, people like, work home, god bless, lot people, wear mask, years ago, virus hoax, like virus, 23 days, grocery store, said virus, 21 million, watch video, 10 days, like amp, uk lockdown | 7.06% |
| 13 | right now, dont want, 3 weeks, tests positive, donald trump, weeks ago, weeks lockdown, virus spreading, coronavirus update, new zealand, 22 million, sounds like, total cases, lockdown 2, communist party, day day, chinese communist, cases 1, whats happening, 2 weeks | 7.18% |

**Sentiment analysis of each latent topic**

We presented the results of the sentiment analysis for each of the thirteen latent topics in Table 3. We also ran a one-tailed z test to examine if each of the eight emotions is statistically significantly different across topics. The *p*-value smaller than .01 was set as a threshold for significance. For example, statistical significance for anticipation in Topic 2 indicated that it was very likely ($p < .001$) that the anticipation emotion is more prevalently expressed in Topic 2 (21.7%) than all other topics.

Table 3. Percentage of each sentiment within 13 topics

| Topic | Anger | Anticipation | Disgust | Fear | joy | sadness | Surprise | Trust |
|---|---|---|---|---|---|---|---|---|
| 1 | 10.80% | 17.60% | 2.00% | 14.60% | 4.60%*** | 2.40% | 1.60% | 9.50% |
| 2 | 12.00% | 21.70%*** | 3.00%*** | 16.90%*** | 4.00% | 4.10%*** | 2.10% | 12.40%*** |
| 3 | 12.60%*** | 17.60% | 2.90%*** | 14.90% | 3.20% | 3.80%*** | 2.60%*** | 15.90%*** |
| 4 | 13.20%*** | 20.90%*** | 3.30%*** | 15.10% | 4.20% | 3.30%*** | 2.20%*** | 13.30%*** |
| 5 | 12.40%*** | 23.80%*** | 2.60%*** | 14.30% | 4.30% | 3.50%*** | 2.10% | 13.40%*** |
| 6 | 13.10%*** | 22.50%*** | 2.40% | 13.40% | 4.60%*** | 3.50%*** | 3.00%*** | 12.80%*** |
| 7 | 12.50%*** | 21.90%*** | 2.50%*** | 17.00%*** | 3.70% | 3.20%*** | 3.30%*** | 13.10%*** |
| 8 | 13.80%*** | 20.70%*** | 2.40% | 16.50%*** | 3.80% | 3.10%*** | 2.40%*** | 12.10%*** |
| 9 | 12.50%*** | 20.70%*** | 2.80%*** | 15.50% | 7.90%*** | 3.40%*** | 2.40%*** | 12.30%*** |
| 10 | 14.60%*** | 17.40% | 3.00%*** | 18.80%*** | 3.30% | 3.30%*** | 1.90% | 11.30% |
| 11 | 11.80% | 20.60%*** | 2.50%*** | 15.50%*** | 6.00%*** | 3.70%*** | 2.70%*** | 11.90%*** |
| 12 | 12.50%*** | 21.40%*** | 2.80%*** | 17.90%*** | 4.20% | 3.30%*** | 2.60%*** | 14.20%*** |
| 13 | 13.30%*** | 20.80%*** | 2.60%*** | 14.80% | 4.30% | 4.20%*** | 3.10%*** | 11.50%*** |

Notes: The sum of the percentage under each topic is not equal to 100%. The rests are either neutral or other emotions.

$p$-value from Z-test, *** $p < .001$

Figure 5 presented eight emotions of trust, anticipation, joy, surprise, anger, fear, disgust, and sadness. Results showed that across all 13 topics, *anticipation* (dark blue line) dominated 12 topics, and followed by *fear* (orange line), *trust* (grey line), and *anger* (yellow line). For example, about 23.8% of Tweets in Topic 8 revealed feelings of anticipation that "necessary steps and precautions will be taken" [18 29]. The emotion *fear* for the impacts of the virus took 18.8% of the Tweets in Topic 10, which was statistically different from the fear expressed in other topics.

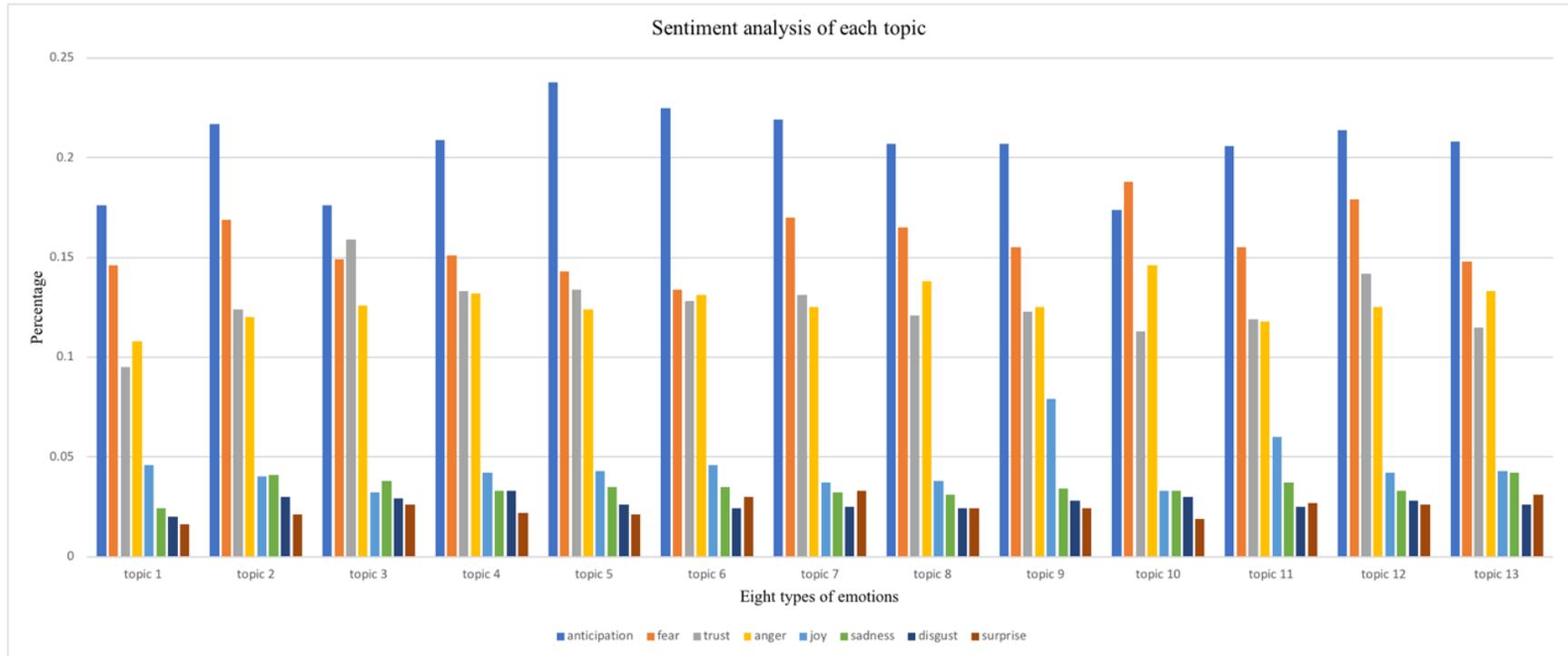

**Figure 6. Sentiment analysis for each of the 13 latent topics**

**COVID-19 related themes**

The qualitative content analysis approach allowed users to categorize these topics into different distinct themes. Two team members discussed these bigrams and generated Tweets samples in each topic and then categorized the identified 13 topics into 5 different themes. To protect the privacy and anonymity of the Twitter users, we did not present any user-related information, such as users' twitter handles or other identifying information; therefore, sample Tweets shown in Table 4 were excerpts drawn from original Tweets.

Table 4. Themes based on topic classification and sample tweets

| Theme | Topic | Common co-occurred words | Sample tweets |
|---|---|---|---|
| Public health Measures to slow the spread of COVID-19 | Facemasks | face masks, wear masks | "we protect us and our family by wearing masks every day." |
| | Quarantine | home quarantine, quarantine period, week quarantine, self quarantine | "@truth_leah @SusanMayGudge @realDonaldTrump @JustineTrudeau They're all under mandatory 2 week quarantine, and they are essential workers…" |
| | Test kits | test kits, testing kits | "… Hydroxychloroquine, Testing Kits and USA: We urge the Modi govt to draw proper lessons from this latest instance of US…" |
| | Lockdown | covid19 lockdown, lockdown period, UK lockdown, weeks lockdown, | "people are actually shocked the lockdown has been extended for 3 weeks when there are still people going out meeting …" |
| | Safety | stay safe, safe stay, stay away | "Be strong, stay safe #lockdown but not locked out …http://t.co/FvifiEbbs7 .." |
| | a need for the vaccine | coronavirus vaccine | "…lead scientist for NIH working on #coronavirus vaccine research …" |
| | U.S.'s shelter-in-place | Shelter place, shelter in | "Did California's shelter-in-place order work? If you sue crap data without any reference to epidemiology, then yes …" |
| Social stigma associated with COVID-19 | Chinese Communist Party (CPP) and the spread of the Virus | communist party, chinese communist, cases 1 | "The #Chinese Communist Party (#CCP) is spreading disinformation to cover up the origin of the #coronavirus" |
| | Discriminatory names | Wuhan virus, wuhan lab, Chinese virus | "…that China is responsible for putting entire world @great risk. Heavily criticized their eating habits." |
| | President Trump's calling Chinese virus | president trump, social media, china virus, trump said, | "President Trump: They know where it came from. We all know where it came from, #chinesevirus..." |
| Coronavirus new cases and deaths | New cases | new cases, total number, number cases, confirmed cases | "RT @neeratanden: 4,591 people died in a day from the virus, the highest number anywhere ever that we know of." |
| | Deaths | death toll, people died, people dying, coronavirus death, | "#Britain's death toll could be DOUBLE official tally as care homes" |
| COVID-19 in the United States | Mental health, and COVID-19 in New York | new york, shelter place, mental health, coronavirus covid | "…New Yorkers on their apartment roofs during quarantine is a whole different vibe. This is gonna be in history books …" |
| | Protests against the lockdown | anti lockdown, people protesting, protesting stay | "I stand with the Healthcare workers!!! Bravo! Healthcare workers face off against anti-lockdown protesters in Colorado …" "…nurses blocking anti lockdown protests in Denver …" |
| | Task force in the US | task force | "RT @Jim_Jordan: There are #coronavirus task forces doing great work. But there is one task force that's missing in action: the U.S. congress" |
| | Coronavirus pandemic in the United States | united states, u s, white house, new york, new jersey, 21 million, million people, dr fauci, | "stay-at-home orders continue in much of the United States …" |
| Coronavirus cases in the rest of the world | UK | Herd immunity, UK lockdown, Prime Minister | "The Prime Minister gave the game away early on when he openly said to Scrofulous and Willibooby that the governments plan was Herd Immunity the REAL people in charge must have been so furious with hinm he had to be sent to an isolation ward with the virus to shut him up!" |
| | Global issue | Entire world, south korea, world health, global pandemic, new zealand | "Worldwide it is now 182,726.." "New Zealand Prime Minster Jacinda Ardern says the government will partially relax its lockdown in a week, as a decline in …" |

We organized thirteen topics into five themes, including "public health measures to slow the spread of COVID-19 (e.g., facemasks, test kits, vaccine)," "social stigma associated with COVID-19 (e.g., Chinese virus, Wuhan virus)," "coronavirus news cases and deaths (e.g., new cases, deaths)," "COVID-19 in the United States (e.g., New York, protests, task force)," and "coronavirus cases in the rest of the world (e.g., UK, global issue)." For example, theme "public health measures to slow the spread of COVID-19" included relevant topics of "facemasks," "quarantine," "test kits," "lockdown," "safety," "a need for the vaccine," and "U.S.'s shelter-in-place." In addition, "home quarantine," "self-quarantine" were two of the most commonly co-occurred words under topic quarantine.

## Discussion

**Principal results**

In this study, we address the problem of public discussions and emotions using COVID-19 related messages on Twitter. Twitter users discuss five main themes related to COVID-19 Between March 7 to April 21, 2020. Topic modeling of the Tweets is effective in providing insights about coronavirus topics and concerns on Twitter. The results show several essential points. First, the public is using a variety of terms referring to COVID-19, including "virus," "COVID 19," "coronavirus," "corona virus." In addition, coronavirus has been referred to as the "China virus" that can create stigma and harm efforts to address the COVID-19 outbreak [14]. Second, discussions about the pandemic in New York are salient, and its associated public sentiments are *anger*. Third, public discussions about the Chinese Communist Party (PPC) and the spread of the virus emerge as a new topic, which is not identified in a previous study [18], suggesting the connection between the COVID-19 and politics is increasing to be circulating on Twitter as the situation evolves. Fourth, public sentiment on the spread of coronavirus is *anticipation* for the potential measures that can be taken and followed by a mixed feeling of trust, anger, and fear. Results suggest that the public is not surprised by the rapid spread of growth. Fifth, the public reveals a significant feeling of *fear* when they discuss the coronavirus crisis and deaths. Lastly, results show that trust for the authorities no longer remain as a prominent emotion when Twitter users discuss COVID-19, which is different than an early study [18] showing that people hold trust for public health authorities based on a sample of Tweets posted from January 20 to March 7, 2020.

**Comparison with prior work**

Our findings are consistent with previous studies using social media data to assess the public health response and sentiments for COVID-19, and suggest that public attention has been focusing on the following topics since January 2020, including

(1) the confirmed cases and death rates [11, 18, 30];

(2) preventive measures [11, 18, 30];

(3) health authorities and government policies [10, 18];

(4) an outbreak in New York [18];

(5) COVID-19 stigma by referencing the coronavirus as the "Chinese virus" [11, 14].

(6) negative psychological reactions (e.g., fear) or mental health consequences [11, 31, 32]

Compared with the study examining public discussions and concerns for COVID-19 using Twitter data from January 20 to March 7, 2020, we find that several salient topics are no longer popular, including (1) Outbreak in South Korea; (2) Diamond princess cruise; (3) economic impact [11 33]; and (4) supply chains [18]. Given the preventive measures, washing hands is no longer a prevalent topic. Instead, quarantine has become dominant.

In addition, our study identifies new discussion topics around the COVID-19 between March 7 to April 21, such as (1) a need for a vaccine to stop the spread; (2) quarantine and shelter-in-place order; (3) pretests against the lockdown; (4) coronavirus pandemic in the United States. The new salient topics suggest that Twitter users (English) are focusing their attention to the COVID-19 in the United States (e.g., New York, protests, task force, millions of confirmed cases) rather than discussions about global news contents (South Korea, Diamond princess cruise, Dr. Li Wenliang in China).

**Limitations**

First, we only sample a trending of 20 hashtags as the key search terms to collect Twitter data (Appendix 1). New hashtags keep coming up as the situation evolves. For example, a hashtag may become widely used after a related topic becomes more popular, such as the official name, COVID-19, for the virus. Second, Twitter users are not representative of the whole population globally, and topics of Tweets only indicate online users' opinions about and reactions to COVID-19. However, the Twitter dataset is still a valuable source allowing us to examine real-time Twitter users' responses and online activities related to COVID-19. Third, non-English Tweets are removed from our analyses, and hence the results are limited to users who posted in English only. Future studies are recommended to include other languages, such as Italian, French, Germany, and Spanish, for COVID-19 analyses.

**Future research**

First, future research could further explore public trust and confidence in existing measures and policies, which is essential. Compared to prior work, our study shows that Twitter users reveal a feeling of *joy* when talking about "herd immunity." We also find sentiments of *fear* and *anticipation* related to topics of quarantine and shelter-in-place. In addition, future studies could evaluate how government officials (e.g., President Trump) and international organizations (e.g., WHO) deliver and convey messages to the public, and its impact on the public opinions and sentiments. Third, future studies could examine the spread of anti-Chinese/Asian sentiments social media and how people use social media platforms to resist and challenge COVID-19 stigma. Fourth, our findings do not show that misinformation during the COVID-19 pandemic is a prominent theme. An existing study shows that 25% (n=153) of sampled Tweets contained misinformation [35]. They also find that the term "COVID-19" has lower rates of misinformation than that associated with the terms of "#2019_ncov" and "Corona." Future research is suggested in the investigation of misinformation and how it expands on social media. Finally, the study finds that trust is no longer prominent when people tweet about confirmed cases and deaths. Instead, fear has replaced trust to be the dominant emotion. Future research is suggested to examine the changes in trust over time.

**Conclusions**

The study shows that Twitter data and machine learning approaches can be leveraged for infodemiology study by studying the evolving public discussions and sentiments during the COVID-19 pandemic. Our findings facilitate an understanding of public discussions and concerns about COVID-19 pandemic among Twitter users between March 7 and April 21, 2020. Several topics were consistently dominant on Twitter, such as "the confirmed cases and death rates," "preventive measures," "health authorities and government policies," "COVID-19 stigma," and "negative psychological reactions (e.g., fear)." As the situation evolves rapidly, new salient topics emerge accordingly. *Anticipation* for the potential measures that can be taken to stop the spread of COVID-19 is still significantly prevalent across all topics. However, Twitter users reveal *fear* when tweeting about COVID-19 new cases or death rather than trust [18]. Real-time monitoring and assessment of the Twitter discussion and concerns can be promising for public health emergency responses and planning. Hearing and reacting to real concerns from the public can enhance trust between the healthcare systems and the public as well as prepare for a future public health emergency.

Appendix

Appendix 1. The list of hashtags used as search terms for data collection

| #COVID19 | #Covid19 | #covid19 | #Covid_19 | #COVID |
| #coronavirus | #Coronavirus | #CoronaVirus | #2019nCoV | #CoronavirusOutbreak |
| #StayHome | #stayhome | #Lockdown | #lockdown | #CoronavirusPandemic |
| #Qurantine | #qurantine | #2019nCoV | #FireTrump | #SARsCov2 |